\documentclass[aps,prd,showpacs,floatfix,nofootinbib,12pt]{revtex4}

\usepackage{color,amsmath,amssymb,graphicx,latexsym,subfigure}
\usepackage{threeparttable}
\usepackage[dvipsnames]{xcolor}
\usepackage{hyperref}

\def\likedm{\textsc{LikeDM}}

\newcommand{\sigmav}{\ensuremath{\langle\sigma v\rangle}}

\usepackage{listings}
\usepackage{color}

\definecolor{dkgreen}{rgb}{0,0.6,0}
\definecolor{gray}{rgb}{0.5,0.5,0.5}
\definecolor{mauve}{rgb}{0.58,0,0.82}

\lstdefinestyle{BASHstyle}{frame=tb,
backgroundcolor=\color{gray!20},
  language=sh,
  aboveskip=3mm,
  belowskip=3mm,
  showstringspaces=false,
  columns=flexible,
  basicstyle={\scriptsize\ttfamily},
  numbers=none,
  numberstyle=\tiny\color{gray},
  keywordstyle=\color{blue},
  commentstyle=\color{dkgreen},
  stringstyle=\color{mauve},
  breaklines=true,
  breakatwhitespace=true,
  tabsize=3
}
\lstdefinestyle{F95style}{frame=tb,
backgroundcolor=\color{yellow!30},
  language=[95]Fortran,
  aboveskip=3mm,
  belowskip=2mm,
  showstringspaces=false,
  columns=flexible,
  basicstyle={\scriptsize\ttfamily},
  numbers=none,
  numberstyle=\tiny\color{gray},
  keywordstyle=\color{blue},
  commentstyle=\color{dkgreen},
  stringstyle=\color{mauve},
  breaklines=true,
  breakatwhitespace=true,
  tabsize=1
}

\lstnewenvironment{BASH}{\pagebreak[2]\lstset{style=BASHstyle}}{\pagebreak[2]}
\lstnewenvironment{F95}{\pagebreak[2]\lstset{style=F95style}}{\pagebreak[2]}

\begin{document}

{\small
\begin{flushright}
IPMU16-0037 \\
TUM-HEP-1039-16
\end{flushright} }

\voffset 1.25cm

\title{\likedm : likelihood calculator of dark matter detection}

\author{Xiaoyuan Huang$^1$\footnote{huangxiaoyuan@gmail.com}}
\author{Yue-Lin Sming Tsai$^2$\footnote{smingtsai@gmail.com}}
\author{Qiang Yuan$^{3,4}$\footnote{yuanq@pmo.ac.cn}}

\affiliation{
$^1$Physik-Department T30d, Technische Universit\"at M\"unchen, James-Franck-Stra\ss{}e, D-85748 Garching, Germany\\
$^2$Kavli IPMU (WPI), University of Tokyo, Kashiwa, Chiba 277-8583, Japan\\
$^3$Department of Astronomy, University of Massachusetts, 710 North
Pleasant St., Amherst, MA, 01003, USA\\
$^4$Key Laboratory of Dark Matter and Space Astronomy, Purple Mountain
Observatory, Chinese Academy of Sciences, Nanjing 210008, China
}

\begin{abstract}
With the large progress in searches for dark matter (DM) particles 
with indirect and direct methods, we develop a numerical tool that 
enables fast calculations of the likelihoods of specified DM particle
models given a number of observational data, such as charged cosmic rays
from space-borne experiments (e.g., PAMELA, AMS-02), $\gamma$-rays from the
Fermi space telescope, and underground direct detection experiments.
The purpose of this tool --- \likedm, likelihood calculator for dark matter
detection --- is to bridge the gap between a particle model of DM and the observational
data. The intermediate steps between these two, including the astrophysical
backgrounds, the propagation of charged particles, the analysis of Fermi
$\gamma$-ray data, as well as the DM velocity distribution and the nuclear 
form factor, have been dealt with in the code. We release the first 
version (v1.0) focusing on the constraints from indirect detection of DM with charged cosmic and gamma 
rays. Direct detection will be implemented in the next version.
This manual describes the framework, usage, and related physics of the code.
\end{abstract}
\date{\today}

%95.35.+d: Dark matter
%96.50.S-: Cosmic rays

\maketitle

{\bf PROGRAM SUMMARY}
  %Delete as appropriate.

\begin{small}
\noindent
{\em Program Title: \likedm}                                          \\
{\em Licensing provisions: GPLv3}                                   \\
{\em Programming language: FORTRAN 90 and Python}                                   \\
%{\em Supplementary material:}                                 \\
  % Fill in if necessary, otherwise leave out.
%{\em Journal reference of previous version:}                  \\
  %Only required for a New Version summary, otherwise leave out.
%{\em Does the new version supersede the previous version?:}   \\
  %Only required for a New Version summary, otherwise leave out.
%{\em Reasons for the new version:}\\
  %Only required for a New Version summary, otherwise leave out.
%{\em Summary of revisions:}*\\
  %Only required for a New Version summary, otherwise leave out.
{\em Operating system: Linux.}\\
{\em Nature of problem: Dealing with the intermediate steps between a dark matter model and data.}\\
{\em Solution method: Fast computation of the likelihood of a given dark matter model (defined by a mass, cross section or decay rate, and annihilation or decay yield spectrum), without digging into the details of cosmic-ray propagation, Fermi-LAT data analysis, or related astrophysical backgrounds.}\\
%{\em Additional comments including Restrictions and Unusual features (approx. 50-250 words):}\\
  %Provide any additional comments here.
   \\

\end{small}

%%#######################################################%%
\section{Introduction \label{sec:intro}}
%%#######################################################%%

After the discovery of the $125$ GeV Higgs boson at the Large Hadron Collider
\cite{ATLAS,CMS}, we have a complete picture of the standard model 
(SM) of particle physics. The next step beyond the SM could be the
identification of the dark matter (DM) particles that were suggested to
be widely present in the Universe by a series of astronomical observations.
Although the astronomical evidence could be attributed to gravitational 
interactions between DM and SM particles, we are yet to exclude 
the possibility of weakly interacting massive particles (WIMPs).
The potential weak interactions between DM and SM particles provide
us with the opportunity to identify DM, directly via  collisions between
DM particles and underground targets or indirectly via the products of DM 
annihilation or decay in the Universe. Many efforts have been made to
find a direct signal of DM in an underground detector;
however, no convincing evidence has been found till date~\cite{Aprile:2012nq,
Akerib:2013tjd,Yue:2014qdu,Xiao:2014xyn}. On the other hand, with the
operation of several new-generation space telescopes and detectors, such
as PAMELA, AMS-02, and Fermi, many anomalies have been found in the
high-energy sky \cite{Cirelli:2012tf,Bergstrom:2012fi,Bi:2014hpa}.
The uncertainties from astrophysical backgrounds and/or astrophysical
sources, however, make the identification of possible DM signals more
challenging. Nevertheless, the constraints on DM models have become more and
more stringent with the new direct and indirect data. Some of
these constraints depend on certain assumptions about the backgrounds
(e.g., the positron anomaly \cite{Adriani:2008zr,Aguilar:2013qda}).
Since there is no consistent signal of DM present in all observations,
we may expect that the assumptions of astrophysical contributions to those
anomalies are reasonable. The combination of various kinds of observations
is expected to give much improved constraints on DM models, which is
one of the motivations for developing this tool for calculating the DM likelihood.

Another motivation is that it is non-trivial to confront DM models
with observational data due to the complicated astrophysical backgrounds.
First, a proper modeling of the backgrounds, with possible systematic 
uncertainties (e.g., the cosmic ray (CR) propagation parameters), is 
necessary when calculating the likelihood of a DM signal. Second, it is
better to decouple the DM model inputs from the following astrophysical 
processes, as it enables our tool to be applied to any DM particle model. 
Third, we intend to have an efficient computation of the DM signal as well 
as the backgrounds. With these goals, we develop this likelihood 
calculator of DM detection, \likedm. The basic function of \likedm\ is 
to deal with the intermediate steps between a DM model and data. 
To achieve this goal, we 1) calculate the propagation of CR
electrons/positrons and antiprotons with Green's functions with respect to 
energy (e.g., integrated with space and time), 2) model the CR backgrounds
with phenomenological forms, 3) model the $\gamma$-ray emission with
standard Fermi-LAT diffuse emission templates and point sources, and
4) calculate the likelihood map of $\gamma$-rays on the ``energy-flux''
plane for given regions of interest (ROIs). Some works have been published 
based on parts of these methods \cite{Tsai:2012cs,Arhrib:2013ela}.
Here we present the first version of this tool and make the code publicly available in
the community and summarize the details in this manual. 
Constraints from direct detection have not been included in this release, 
and will be added in the subsequent version.

This manual is structured as follows. In Sec.~\ref{sec:charge}, 
we describe the calculation of charged CRs from both the DM signal and 
the background. The Green's function for fast computation of the propagation
of charged CRs is presented. In Sec.~\ref{sec:gamma}, we describe 
the likelihood calculation from Fermi-LAT observations of dwarf spheroids 
(dSphs). We give the energy-flux likelihood map with updated 
Fermi-LAT data. We introduce the code, installation procedure, and 
explain the usage of \likedm\ in Sec.~\ref{sec:manual}. 
Finally, we summarize in Sec.~\ref{sec:sum}.

%%#######################################################%%

%%#######################################################%%
\section{Charged cosmic rays \label{sec:charge}}
%%#######################################################%%

\subsection{Propagation of charged cosmic ray particles}

The charged cosmic rays (CRs) propagate diffusively in the random
magnetic field of the Milky Way. The interaction with the interstellar
medium (ISM) will result in energy losses and/or fragmentation of the 
primary CRs, as well as the production of secondary CRs. For 
electrons/positrons, there will be additional energy losses due to 
radiation in the interstellar radiation field (ISRF) and the 
magnetic field. The random shocks in the interstellar space may
reaccelerate the low-energy CR particles. There may also be 
convective transport of CRs as evidenced by the wide existence
of galactic winds. The general propagation equation of CRs in the
Milky Way can be written as \cite{Strong:2007nh}
\begin{eqnarray}
\frac{\partial \psi}{\partial t}
&=& Q({\bf x},p)+\nabla\cdot(D_{xx}\nabla \psi-{\bf
V_c}\psi)+\frac{\partial}{\partial p}p^2D_{pp}\frac{\partial}
{\partial p}\frac{1}{p^2}\psi \nonumber \\
&-& \frac{\partial}{\partial p}
\left[\dot{p}\psi-\frac{p}{3}(\nabla\cdot{\bf V_c}\psi)\right]-
\frac{\psi}{\tau_f}-\frac{\psi}{\tau_r}, \label{prop}
\end{eqnarray}
where $\psi$ is the CR differential density per unit momentum interval, 
$Q({\bf x},p)$ is the source function, $D_{xx}$ is the spatial diffusion 
coefficient, ${\bf V_c}$ is the convection velocity, $D_{pp}$ is the 
diffusion coefficient in momentum space, $\dot{p}\equiv{\rm d}p/{\rm d}t$
is the momentum loss rate, and $\tau_f$ and $\tau_r$ are timescales for
fragmentation and radioactive decay, respectively. A homogeneous spatial 
diffusion coefficient $D_{xx}$ is assumed, and the rigidity dependence 
is assumed to be of a power-law form $D_{xx}=D_0\beta(R/R_0)^{\delta}$, 
with $\beta$ being the velocity of the particle and $\delta$ reflecting the 
property of the ISM turbulence. For Kolmogrov turbulence, 
we have $\delta=1/3$. The reacceleration is described by diffusion 
in momentum space. The momentum diffusion coefficient $D_{pp}$ can be 
related to the spatial diffusion coefficient $D_{xx}$ by \cite{Seo:1994}
\begin{equation}
D_{pp}D_{xx}=\frac{4p^2v_A^2}{3\delta(4-\delta^2)(4-\delta)w},
\end{equation}
where $v_A$ is the Alfven speed, and $w$ characterizes the level of 
turbulence which can be absorbed in $v_A$. The CRs are assumed to be
confined in an extended halo with characteristic height $z_h$, beyond 
which free escape is assumed. Thus, the major propagation parameters are 
$D_0$, $\delta$, $v_A$, $V_c$ and $z_h$. 

The secondary-to-primary ratios, such as B/C and (Sc+Ti+V)/Fe, and the
unstable-to-stable ratios of secondary particles, such as $^{10}$Be/$^9$Be
and $^{26}$Al/$^{27}$Al are often used to determine the propagation 
parameters \cite{Seo:1994,Strong:1998pw,Maurin:2001sj,Putze:2010zn}.
There are numerical codes to compute CR propagation in the galaxy, 
such as GALPROP\footnote{\url{http://galprop.stanford.edu/}} \cite{Strong:1998pw} 
and DRAGON\footnote{\url{http://www.dragonproject.org/Home.html}}
\cite{Evoli:2008dv}. 

In this tool, we adopt GALPROP version 50\footnote{For the recent updated
version 54 please refer to 
\url{http://sourceforge.net/projects/galprop/}.} 
to calculate the propagation of charged particles. We adopt six sets of
propagation parameters, with $z_h$ varying from $2$ kpc to $15$ kpc, which
reflect the major uncertainties in the propagation parameters 
\cite{Ackermann:2012rg}. All groups are consistent with 
the B/C data as well as the Fermi diffuse $\gamma$-ray emission data
\cite{Ackermann:2012rr}.

\begin{table}[!htb]
\caption {Propagation parameters.}
\begin{tabular}{ccccc}
\hline \hline
 & $D_0^a$ & $z_h$ & $v_A$ & $\delta$ \\
 & ($10^{28}$cm$^2$\,s$^{-1}$) & (kpc) & (km s$^{-1}$) & \\
\hline
1 & $2.7$ & $2$ & $35.0$ & $0.33$ \\
2 & $5.3$ & $4$ & $33.5$ & $0.33$ \\
3 & $7.1$ & $6$ & $31.1$ & $0.33$ \\
4 & $8.3$ & $8$ & $29.5$ & $0.33$ \\
5 & $9.4$ & $10$ & $28.6$ & $0.33$ \\
6 & $10.0$ & $15$ & $26.3$ & $0.33$ \\
\hline
\hline
\end{tabular}\vspace{3mm}\\
$^a$Diffusion coefficient at $R=4$ GV.\\
\label{table:prop}
\end{table}

\subsection{Green's function of charged particle fluxes from DM}

The annihilation or decay of DM particles in the Milky Way halo will produce 
charged CRs such as positrons and antiprotons, which will experience 
diffusive propagation before reaching the Earth. The fluxes of the charged 
CRs depend on both the density profile of DM and the propagation parameters 
(especially the height of the propagation halo $z_h$). We will consider 
several common forms of DM density profile, including the 
Navarro-Frenk-White (NFW) profile \cite{Navarro:1996gj}
\begin{equation}
\rho_{_{\rm NFW}}(r)=\frac{\rho_s}{(r/r_s)(1+r/r_s)^2},
\end{equation}
the Einasto (EIN) profile \cite{Einasto:1965}
\begin{equation}
\rho_{_{\rm EIN}}(r)=\rho_s\cdot\left[-\frac{2}{\alpha}\left(\left(
\frac{r}{r_s}\right)-1\right)\right],
\end{equation}
and the isothermal (ISO) profile \cite{Bahcall:1980fb}
\begin{equation}
\rho_{_{\rm ISO}}(r)=\frac{\rho_s}{1+(r/r_s)^2}.
\end{equation}
The profile parameters are given in Table \ref{table:profile} 
\cite{Bertone:2008xr}.

The source function of the charged CRs for DM annihilation or decay is
\begin{equation}
q(E,r)=\left\{\begin{array}{ll}
\frac{\sigmav}{2m_{\chi}^2}\frac{dN}{dE}\times\rho^2(r) & \text{for annihilation}\\
\frac{1}{m_{\chi}\tau}\frac{dN}{dE}\times\rho(r) & \text{for decay}
\end{array}\right.,
\label{source}
\end{equation}
where $m_{\chi}$ is the mass of the DM particle, $\sigmav$ is
the annihilation cross section, $\tau$ is the decay lifetime, and $dN/dE$ is the particle
yield spectrum per annihilation or decay. 

\begin{table}[!htb]
\caption {DM density profile parameters.}
\begin{tabular}{cccc}
\hline \hline
 & $r_s$ & $\rho_s$ & $\alpha$ \\
 & (kpc) & (GeV cm$^{-3}$) & \\
\hline
NFW & $20$ & $0.26$ & N/A  \\
EIN & $20$ & $0.06$ & 0.17 \\
ISO & $5$  & $1.16$ & N/A \\
\hline
\hline
\end{tabular}
\label{table:profile}
\end{table}

\begin{figure*}[!htb]
\includegraphics[width=0.45\textwidth]{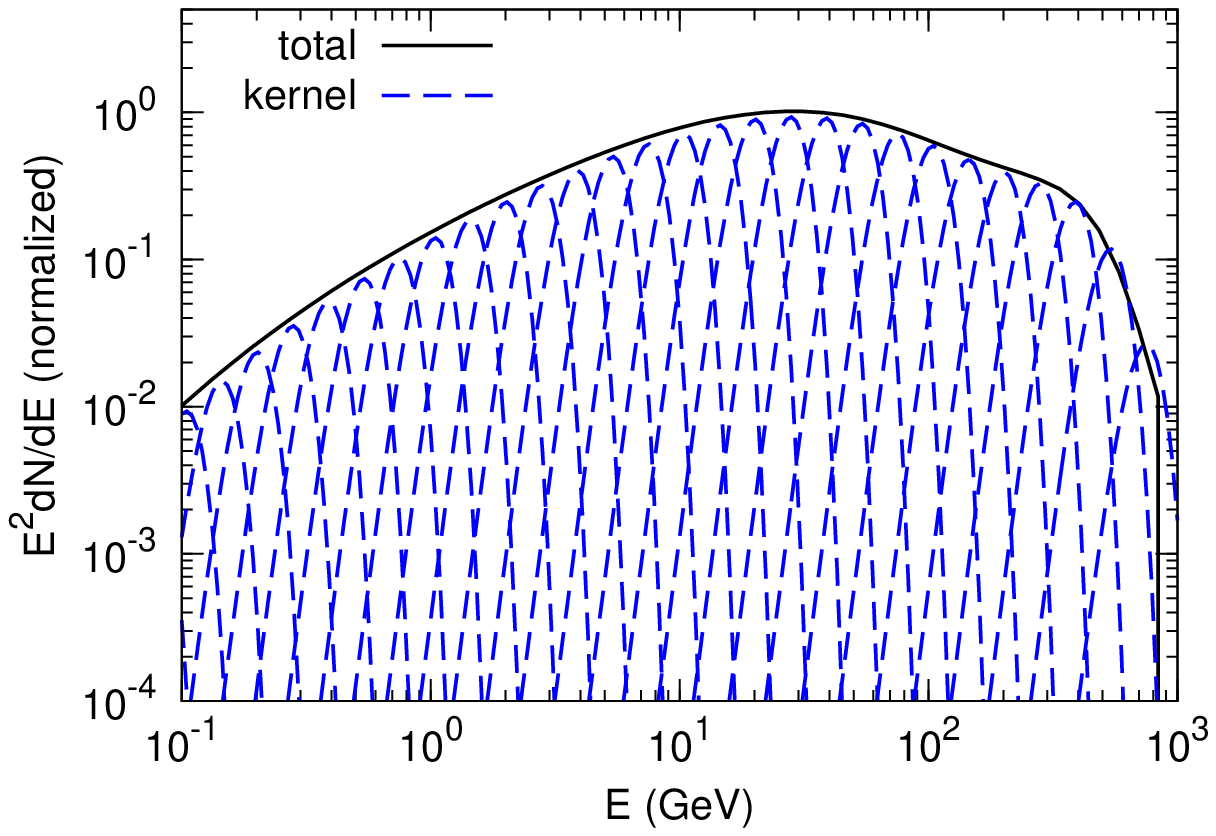}
\includegraphics[width=0.45\textwidth]{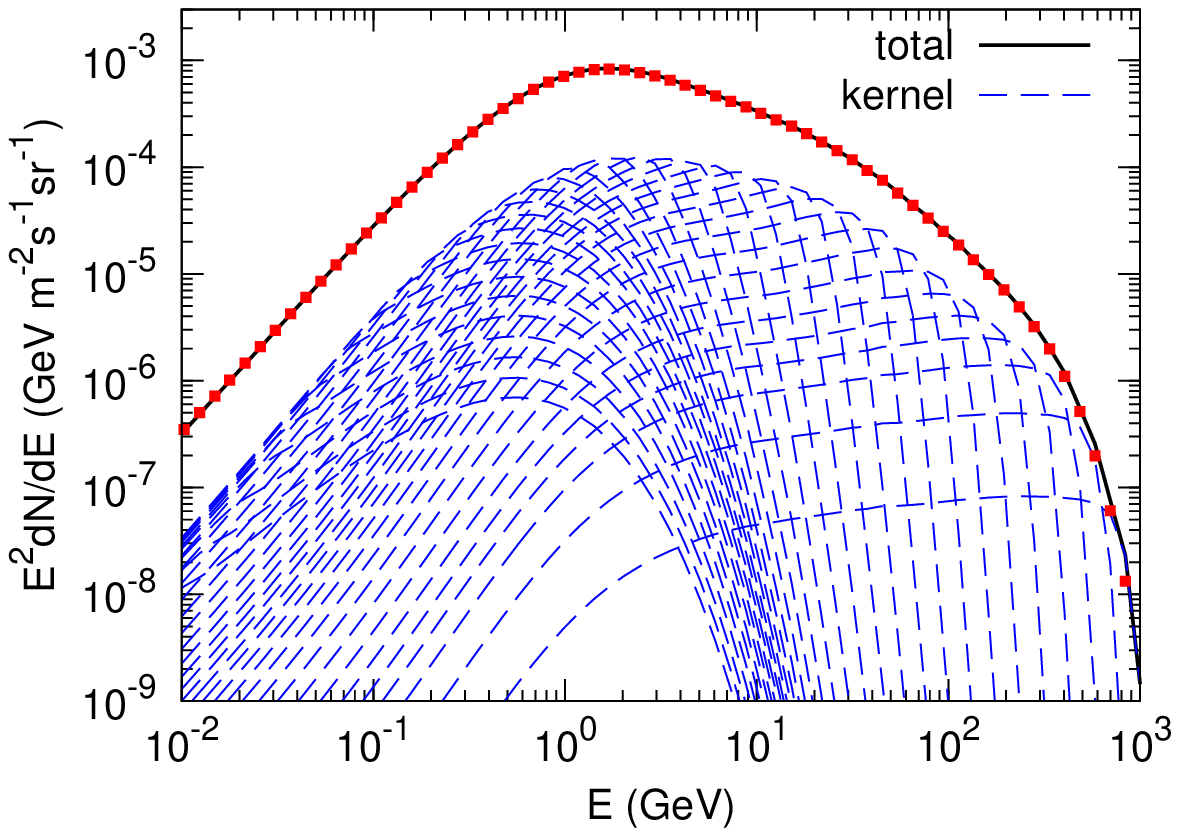}
\caption{Illustration of the kernel functions and the sum of the
positron spectra, before (left) and after (right) the propagation.
The points in the right panel show the direct calculation of the 
propagated spectrum of positrons with GALPROP. Here we adopt the
second setting of the propagation parameters and the NFW profile of
DM density. The mass of DM particle is 1 TeV, with an
annihilation cross section of $\sigmav=10^{-26}$ cm$^3$s$^{-1}$ and $b\bar{b}$ as the
annihilation final state.
\label{fig:prop}}
\end{figure*}

The traditional way to solve the propagation of the DM-induced charged 
particles is to incorporate the source term for a given DM particle model
in the propagation equation (Eq. (\ref{prop})). In order to isolate the
DM particle model from the propagation calculation, we approximate the
function $dN/dE$ with a series of Gaussian kernel functions
\begin{equation}
\frac{dN}{dE}\approx\sum_{i}C_iK_i(E,E_i)=\sum_{i}\frac{C_i}
{\sqrt{2\pi}\sigma_i}\exp\left[-\frac{(E-E_i)^2}{2\sigma_i^2}\right],
\end{equation}
where $E_i$ and $\sigma_i$ are respectively the central value and width of the
$i$th Gaussian kernel. We find that generally $\sigma_i=15\%E_i$
results in a good approximation to most of the energy spectrum $dN/dE$,
except when it has very distinct (e.g., monochromatic) spectral structures.
The left panel of Fig. \ref{fig:prop} shows an illustration of the kernel 
functions weighted by the coefficients $C_i$ for a given spectrum.
We can then calculate the propagated spectrum of each kernel function,
$G_i(E)$, which is the approximate Green's function with respect to 
energy $E$ (dashed lines in the right panel of Fig. \ref{fig:prop}). 
The total propagated spectrum can be obtained as
\begin{equation}
\Phi(E)\approx\sum_{i}C_iG_i(E).
\end{equation}
As shown in the right panel of Fig. \ref{fig:prop}, the result from this
Green's function method is in good agreement with the direct calculation
of the propagation (red dots).

Upon applying this method, any DM-induced CR $e^\pm$ and $\bar{p}$ spectra 
on Earth can easily be obtained by inserting its source shape. This helps 
us to significantly reduce the computation time\footnote{As a rough example,
the computation time of the propagation calculation reduces by a factor 
of $\gtrsim10$ compared with the use of GALPROP.} if the predetermined 
Green's function tables are provided. Users are allowed to generate their 
own tables of Green's functions if necessary.

\subsection{Backgrounds}

The CR backgrounds relevant for the DM searches include the primary
electrons from the CR sources, the secondary positrons and antiprotons
from interactions between the primary CR nuclei and the ISM, as well
as the possible primary sources of positrons from e.g., pulsars
\cite{Shen:1970}. Instead of using the more physical model which considers 
the injection/production and propagation of each type of particle
\cite{Yuan:2013eja,Yuan:2013eba}, we adopt an empirical model to fit the 
locally observed cosmic ray fluxes following Ref. \cite{Aguilar:2013qda}.
This is equivalent to assuming that there is no DM ``signal'' in the current data
and that all the measured events come from CR backgrounds
(see also \cite{Feng:2012gs,Bergstrom:2013jra}). In contrast to Ref. \cite{Aguilar:2013qda}, we assume broken power-law forms to 
describe the fluxes of the primary electrons, secondary positrons/electrons 
and secondary antiprotons, with the purpose of reproducing the wide-band data:
\begin{eqnarray}
\phi_{e^-}&=&C_{e^-}E^{-\gamma_1^{e^-}}\left[1+(E/E_{\rm br}^{e^-})
^{\gamma_2^{e^-}}\right]^{-1},\label{eq:nu_ep}\\
\phi_{e^+}&=&C_{e^+}E^{-\gamma_1^{e^+}}\left[1+(E/E_{\rm br}^{e^+})
^{\gamma_2^{e^+}}\right]^{-1},\label{eq:nu_em}\\
\phi_{\bar{p}}&=&C_{\bar{p}}E^{\gamma_1^{\bar{p}}}\left(1+E/E_{\rm br}
^{\bar{p}}\right)^{-\gamma_2^{\bar{p}}}.\label{eq:nu_ap}.
\end{eqnarray}
Note that the form of antiprotons is slightly different from that of electrons
and positrons in order to improve the fit to the data. The secondary electron 
spectrum is assumed to be the same as the secondary positron spectrum, with 
a normalization factor of $0.6$ as expected from the $pp$ collisions
\cite{Kamae:2006bf}. As for the extra source to reproduce the 
electron/positron excess, a power law with an exponential cut-off 
is assumed
\begin{equation}
\phi_s=C_sE^{-\gamma_s}\exp(-E/E_c).
\end{equation}
Therefore the total fluxes of positrons, electrons, and positrons + electrons are
\begin{eqnarray}
\Phi_{e^+}&=&\phi_{e^+}+\phi_s,\\
\Phi_{e^-}&=&\phi_{e^-}+0.6\phi_{e^+}+\phi_s,\\
\Phi_{e^{\pm}}&=&\phi_{e^-}+1.6\phi_{e^+}+2\phi_s,
\end{eqnarray}
respectively, and the positron fraction is $\Phi_{e^+}/\Phi_{e^{\pm}}$.

\begin{figure*}[!htb]
\includegraphics[width=0.45\textwidth]{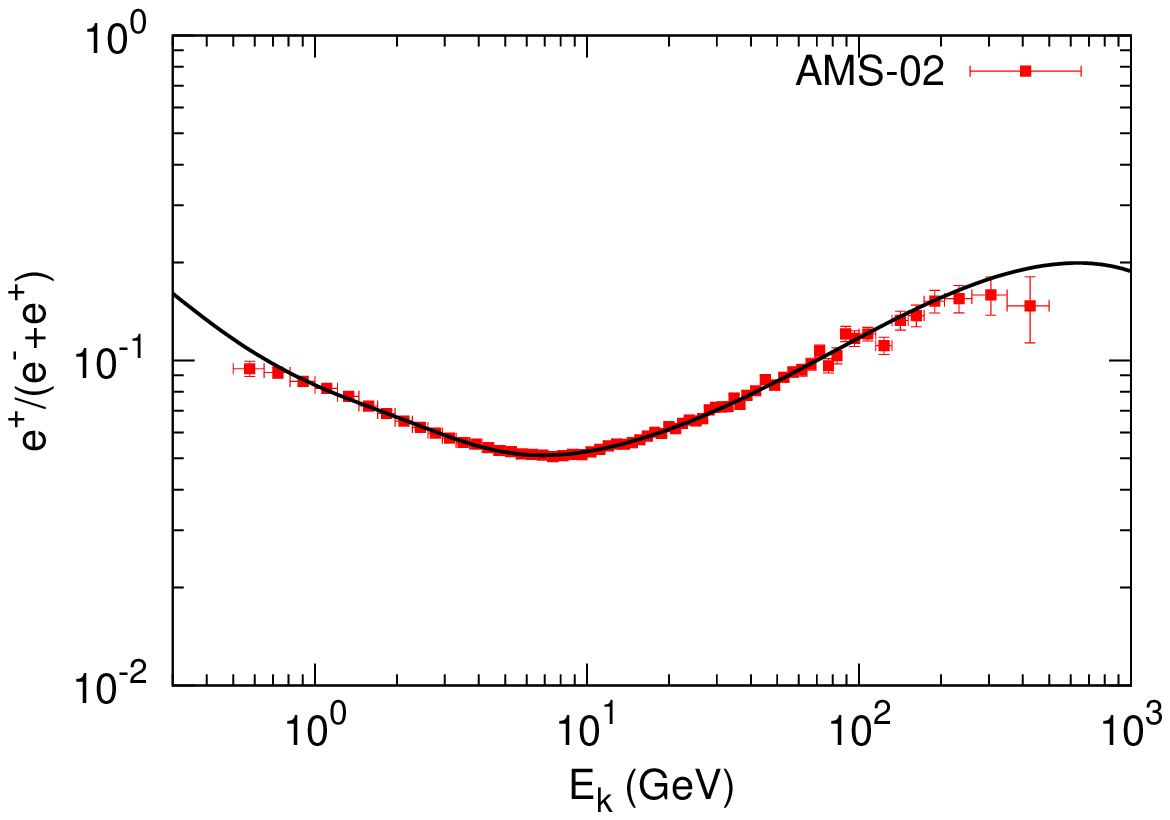}
\includegraphics[width=0.45\textwidth]{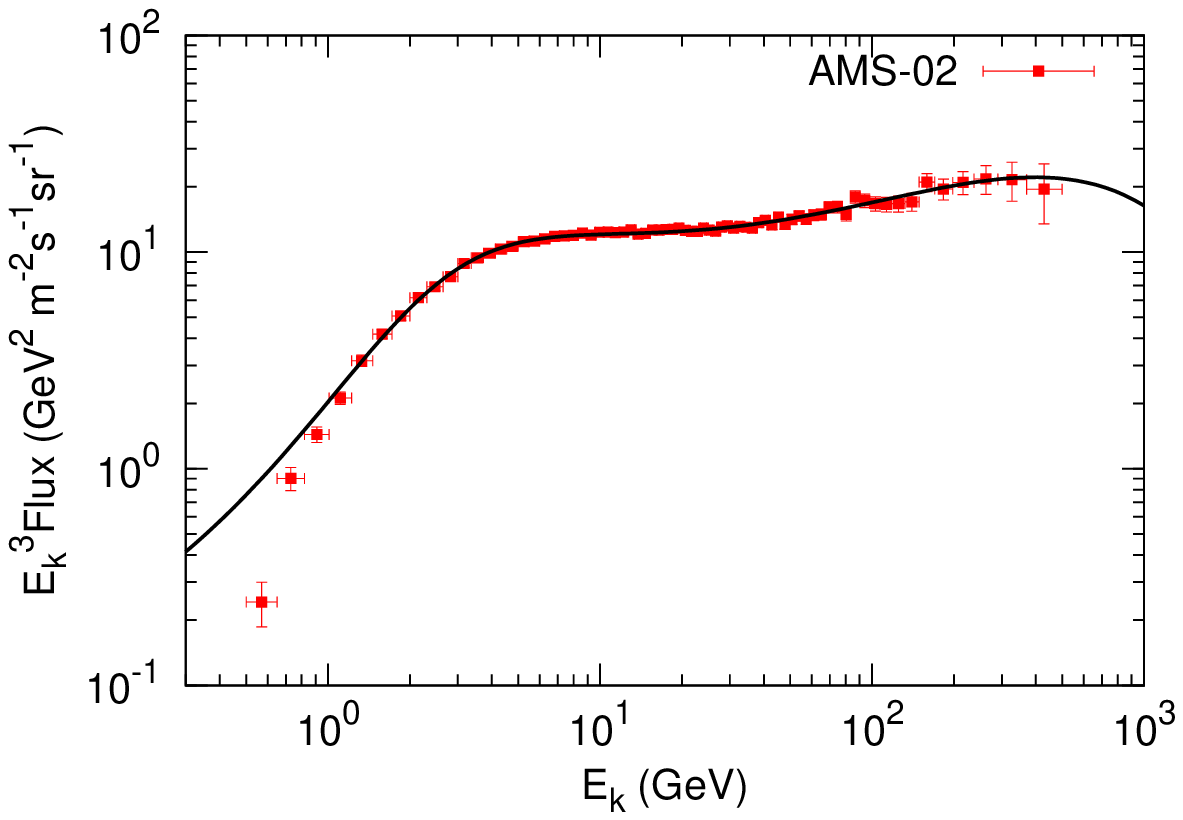}
\includegraphics[width=0.45\textwidth]{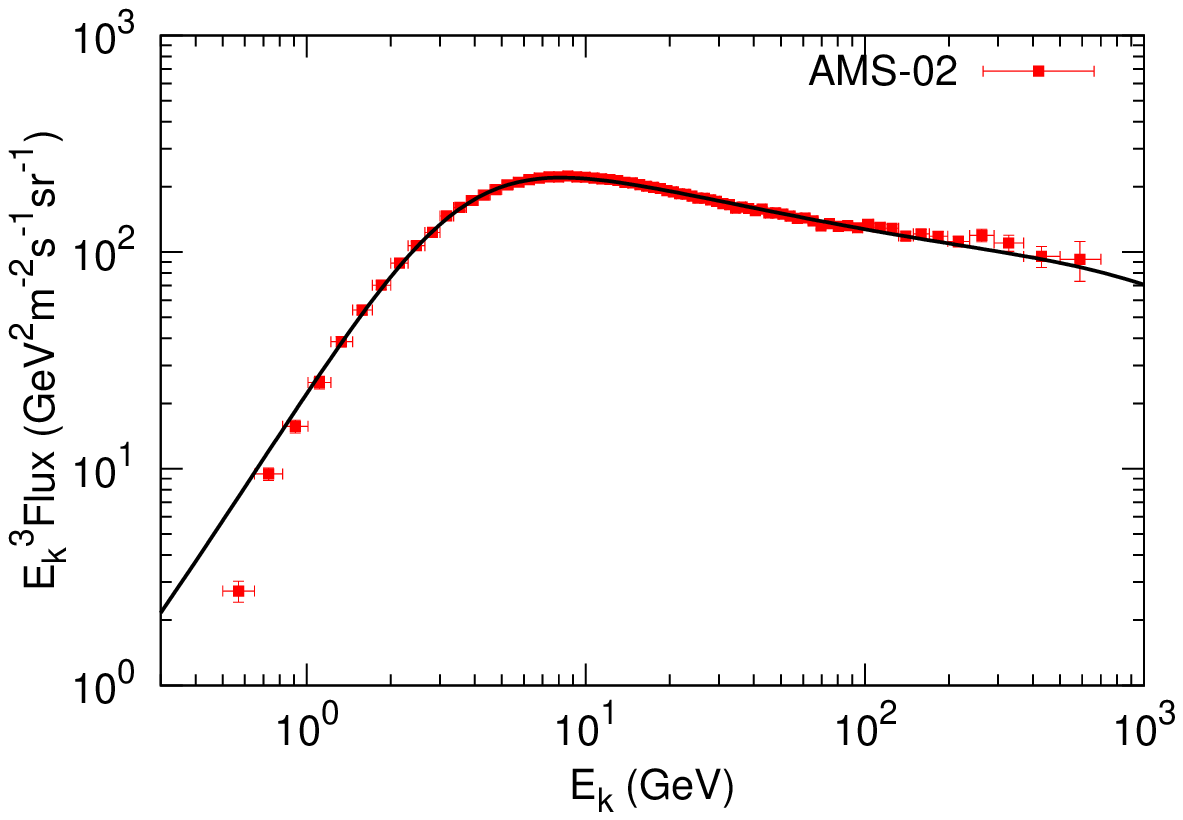}
\includegraphics[width=0.45\textwidth]{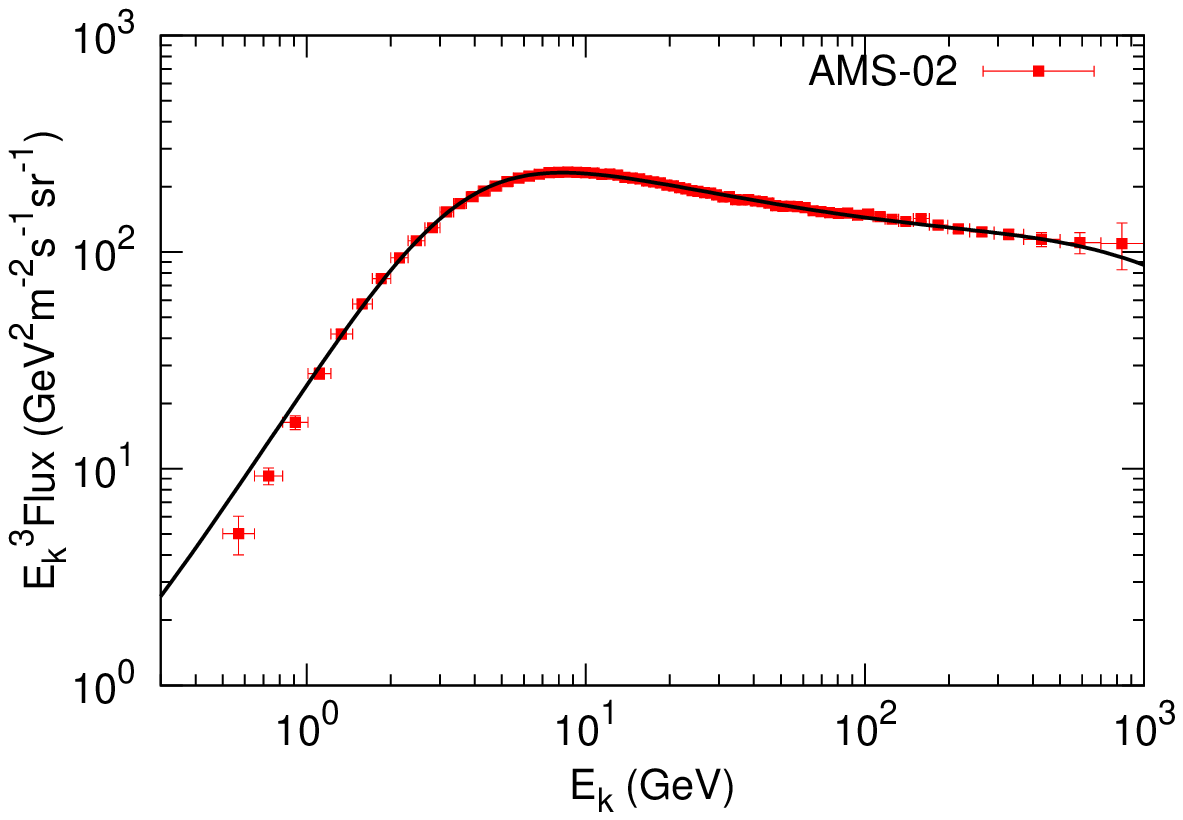}
\includegraphics[width=0.45\textwidth]{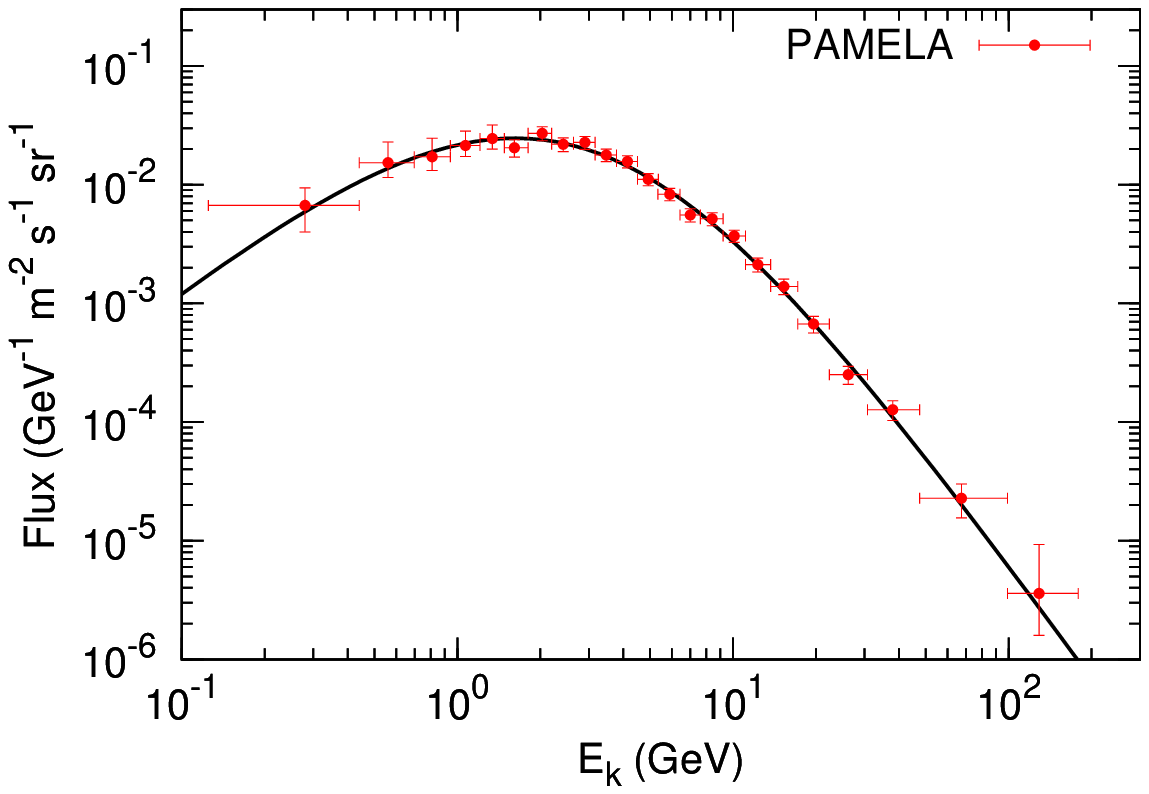}
\caption{Background fitting results of the positron fraction (top-left),
positron (top-right), electron (middle-left), total $e^{\pm}$ (middle-right),
and antiproton spectra (bottom). 
\label{fig:bkg}}
\end{figure*}

The data used to fit the backgrounds includes the updated AMS-02 positron 
fraction \cite{Accardo:2014lma}, the AMS-02 spectra of electrons and 
positrons \cite{Aguilar:2014mma}, the AMS-02 total $e^{\pm}$ spectra 
\cite{Aguilar:2014fea}, and the PAMELA antiproton spectrum 
\cite{Adriani:2010rc}. The AMS-02 data below 1 GeV are excluded from the 
fit \cite{Aguilar:2013qda}. The empirical background model gives a very
good description of the data, as shown in Fig. \ref{fig:bkg}. 
The best-fit $\chi^2$ value over the number of degrees of freedom (dof) 
is about $132.8/285$ for $e^+e^-$ and about $11.1/19$ for antiprotons. 
The best-fit parameters are listed in Table \ref{table:bkg}.

\begin{table}[!htb]
\caption {Best-fit parameters of the backgrounds.}
\begin{tabular}{lccccc}
\hline \hline
 & $C$ & $\gamma_1$ & $\gamma_2$ & $E_{\rm br}$ & $E_c$ \\
 & (GeV$^{-1}$m$^{-2}$s$^{-1}$sr$^{-1}$) &  &  & (GeV) & (GeV) \\
\hline
$\phi_{e^-}$ & $21.6701$ & $0.9344$ & $2.3734$ & $3.6390$ & ... \\
$\phi_{e^+}$ & $1.4991$  & $0.9024$ & $2.3647$ & $2.8434$ & ... \\
$\phi_{s}$   & $0.6526$  & $2.3390$ & ... & ... & $652.89$ \\
$\phi_{\bar{p}}$ & $0.0995$ & $1.844$ & $5.077$ & $2.849$ & ... \\
\hline
\hline
\end{tabular}
\label{table:bkg}
\end{table}

When a DM component is added to the model, we should allow for some freedom
in the backgrounds to obtain a global best-fit to the data. 
Therefore, we multiply by factors of $\alpha_iE^{\beta_i}$, 
with $i=\{e^-,\,e^+,\,s,\,\bar{p}\}$, on the primary electrons, the secondary positrons/electrons,
the extra positrons/electrons and the secondary antiprotons. We adopt
the profile likelihood method to manage the nuisance parameters $\alpha_i$ 
and $\beta_i$, with the scan ranges $[0.1,\,10]$ and $[-0.5,\,0.5]$, 
respectively. The code \texttt{Minuit}~\cite{James:1975dr} is used to 
find the maximum likelihood within the parameter space $[\alpha_i,\beta_i]$.

\subsection{Solar modulation}

The low-energy charged CRs will be modulated by solar activity.
We adopt the simple force-field approximation with only one parameter, \textit{viz.}
the modulation potential, to calculate the effect of solar modulation
\cite{Gleeson:1968zza}. Since our background model is an empirical one
instead of a physical model, the solar modulation only applies to the
CR fluxes from the DM annihilation or decay.

\subsection{The cosmic ray constraints on the DM annihilation parameters space}

In this subsection we present some results on the DM model parameter
constraints from charged CRs derived with the above method. We adopt
a DM annihilation scenario for illustration, and assume that the DM density
profile is NFW. Given one set of the DM model parameters, such as the
mass, the annihilation cross section, and the branching ratios to each 
annihilation channel, we calculate the production spectra of positrons 
and antiprotons using the tables\footnote{
Only those tables of fluxes at production including EW corrections 
are used and incorporated in $\likedm$.}
of Ref.~\cite{Cirelli:2010xx}. 
The propagated fluxes, calculated with the aforementioned Green's 
function method, together with the backgrounds, are then combined with the
data to derive the likelihood, ${\mathcal L}\propto\exp(-\chi^2/2)$, 
of this particular set of DM parameters.

\begin{figure*}[!htb]
\includegraphics[width=0.45\textwidth]{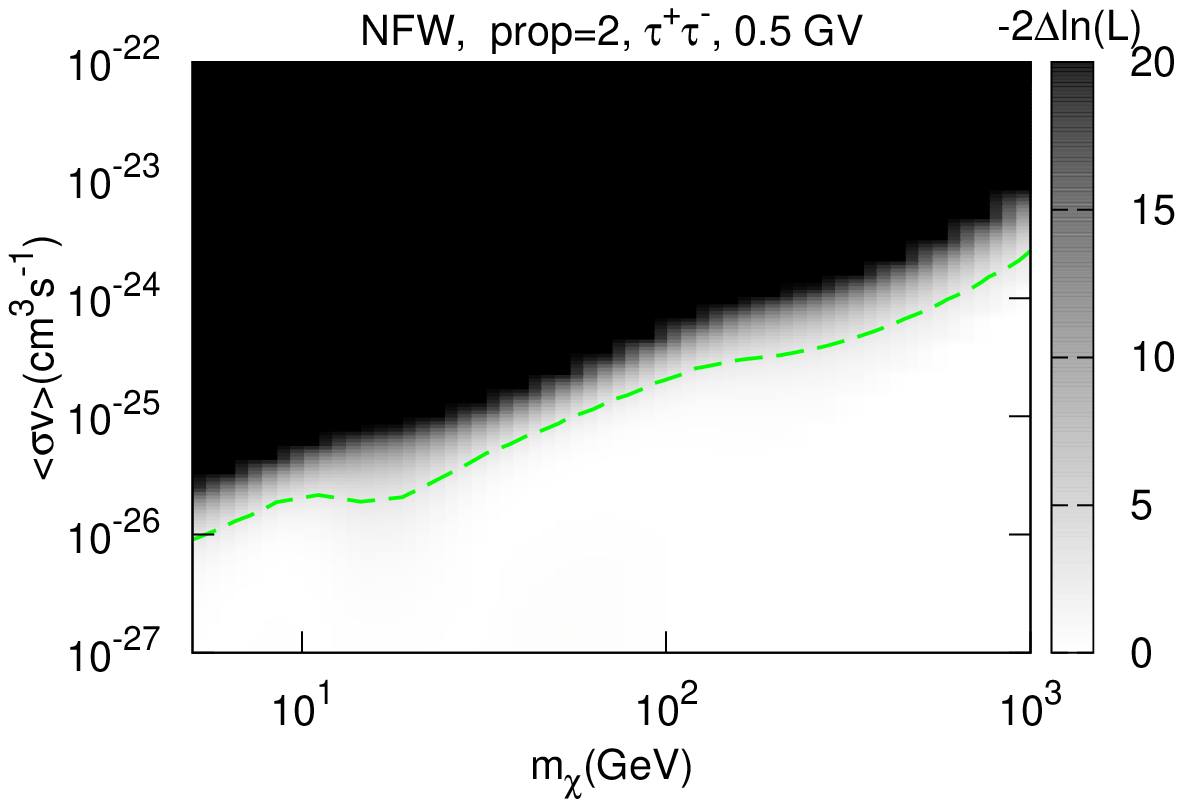}
\includegraphics[width=0.45\textwidth]{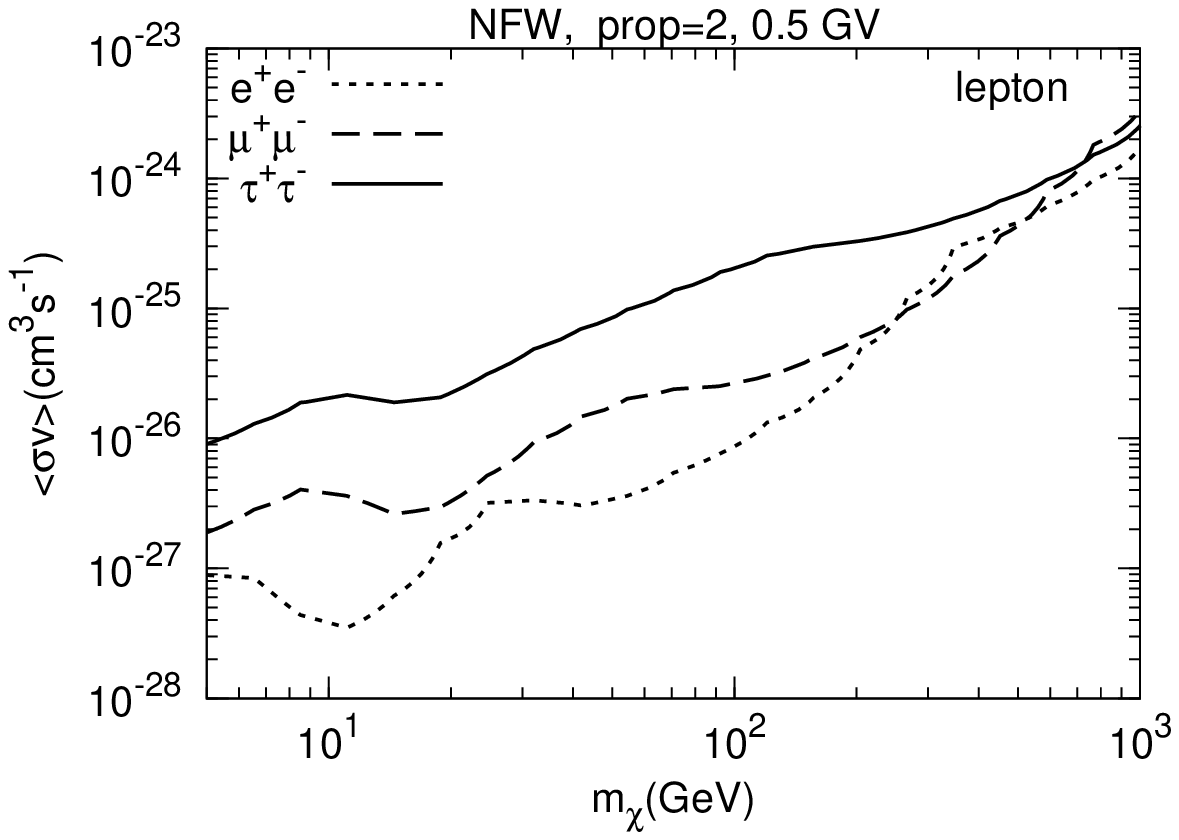}
\includegraphics[width=0.45\textwidth]{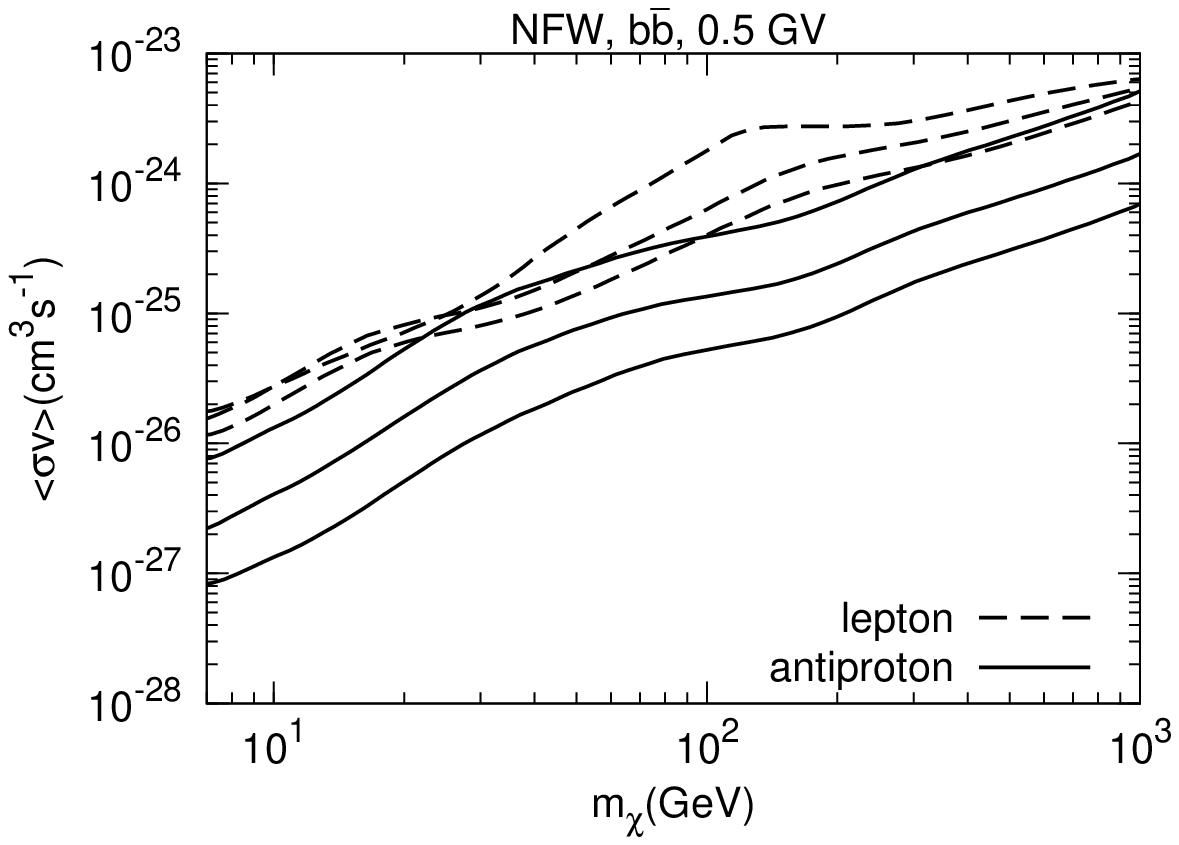}
\includegraphics[width=0.45\textwidth]{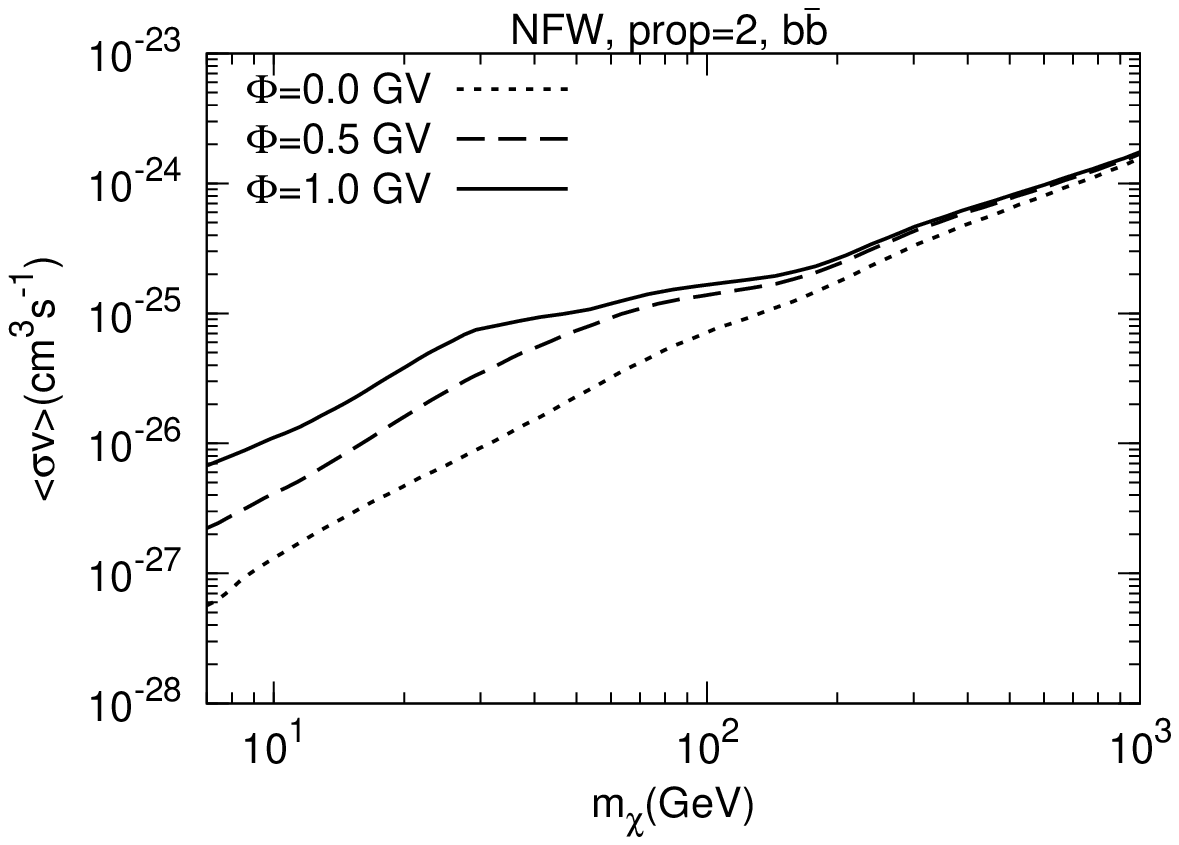}
\caption{Figure shows the constraints on the DM annihilation parameters from the charged CR 
data. The DM density profile is assumed to be NFW, and the solar modulation 
potential is adopted to be 0.5 GV unless stated elsewhere. 
Top-left: the map of $-2\Delta\ln({\mathcal L})$ on the ($m_{\chi},\,\sigmav$)
plane from the AMS-02 lepton data, for DM annihilation into $\tau^+\tau^-$.
The dashed line shows the $95\%$ CL limit. Top-right: $95\%$ upper limits 
of $\sigmav$ from the AMS-02 lepton data, for different DM masses and the 
annihilation channels of $e^+e^-$, $\mu^+\mu^-$, and $\tau^+\tau^-$, 
respectively. Bottom-left: $95\%$ upper limits on $\sigmav$ from the 
AMS-02 lepton data (dashed) and the PAMELA antiproton data (solid), for 
DM annihilation to $b\bar{b}$. The three lines of each group from top to 
bottom represent the propagation models \#1, \#2, and \#6, respectively. 
Bottom-right: $95\%$ upper limits on $\sigmav$ from the combination of 
the AMS-02 lepton data and the PAMELA antiproton data, for DM annihilation 
into $b\bar{b}$. The three lines from top to bottom represent the solar 
modulation potentials of 1.0, 0.5, and 0 GV, respectively.
\label{fig:con}}
\end{figure*}

The top-left panel of Fig. \ref{fig:con} shows a map of $-2\Delta\ln
({\mathcal L})\equiv -2\ln({\mathcal L}/{\mathcal L^0})$, where 
${\mathcal L}$ is the likelihood of the model with different
values of $\sigmav$ and $m_\chi$, and ${\mathcal L^0}$ is 
the likelihood for the null hypothesis (\textit{i.e}., pure background). The likelihood 
is calculated using the AMS-02 $e^+e^-$ data. The propagation model is \#2, 
the solar modulation potential is 0.5 GV, and the DM annihilation channel 
is assumed to be $\tau^+\tau^-$. The dashed line shows the $95\%$ 
confidence level (CL) upper limit, defined by $-2\Delta\ln
({\mathcal L})=2.71$ for a single-sided probability distribution.
Other panels of Fig. \ref{fig:con} illustrate the $95\%$ upper limits
of the DM annihilation cross section for different channels (top-right),
propagation models (bottom-left), and solar modulation potentials
(bottom-right). 

%%#######################################################%%
\section{Gamma-rays from dSphs \label{sec:gamma}}
%%#######################################################%%

Gamma-rays are another very important messenger for the indirect detection 
of DM. Gamma-rays travel through space without deflection, thus they can
point back to the sources emitting them. It is advantageous to choose
regions in the sky with high DM density and low astrophysical background
to search for DM. The dSphs in the Milky Way are widely believed to be 
favorable targets with a high signal-to-noise ratio. Many works have been 
performed to search for DM-induced $\gamma$-rays from dSphs with Fermi-LAT 
data, yet none of them reported a significant detection 
\cite{Ackermann:2011wa,Ackermann:2013yva,Geringer-Sameth:2014qqa,
Ackermann:2015zua,Zhao:2016xie}. Recently, the ongoing Dark Energy Survey 
(DES) reported some new candidates of dSphs in the southern hemisphere 
\cite{Bechtol:2015cbp,Drlica-Wagner:2015ufc}. Several groups had claimed 
possible weak $\gamma$-ray signals from Reticulum 2 
\cite{Geringer-Sameth:2015lua} and Tucana III \cite{Li:2015kag}.
Since there are no reliable kinematic measurements available for these newly-discovered dSphs candidates (hence no reliable DM density profiles), the 
constraints and implications on DM from them are very uncertain. 
We adopt the dSphs sample of Ref. \cite{Ackermann:2015zua} in \likedm.

\subsection{Likelihood Map}

To ensure an easy computation of the total likelihood for any given shape 
of $\gamma$-ray spectrum, we take the likelihood map method first 
proposed in our earlier work~\cite{Tsai:2012cs} and further developed in 
Refs. \cite{Ackermann:2013yva,Ackermann:2015zua}. Briefly speaking, the 
likelihood ${\mathcal L}_{ij}$ of any flux $\phi_j$ in each energy bin 
$[E_{i-1/2},E_{i+1/2}]$ is calculated to give a likelihood map on the 
($E,\,\phi$) plane. The total likelihood of a given spectrum can be simply 
obtained through a product of the likelihoods over all energy bins. This 
method is DM particle model independent, flexible and time-saving. 
Also, as shown in Ref. \cite{Ahnen:2016qkx}, it is simple to combine this method
with data from other observations. 

We describe the method in more detail. DM annihilation in dSphs is adopted 
for illustration\footnote{The decay case is also included in the code. 
Since the emission from DM decay is more extended, we use the profile
parameters, which could get the median J-factor corresponding to 5$^\circ$ 
radius integration, from the Markov Chain Monte Carlo (MCMC) fittings to the 
kinematic data of dSphs \cite{Bonnivard:2015xpq} to generate normalized 
two-dimensional \textbf{SpatialMap}. Then we calculate the likelihood map 
for these extended sources. For J-factors and corresponding errors please see 
Ref.~\cite{web:J_err}.}.
The case of DM decay can be easily obtained via proper
re-adjustment of the formula (see e.g., Eq. (\ref{source})). 
The $\gamma$-ray flux from the annihilation of DM in a dSph is
\begin{equation}
\phi(E)={\frac{\sigmav}{8{\pi}m_{\chi}^{2}}\times \frac{dN_{\gamma}}
{dE_{\gamma}}\times J},
\end{equation}
where $J={\int}dld{\Omega}{\rho}(l)^{2}$ is the so-called $J$-factor which 
characterizes the amount of annihilation from a specified direction given 
the density distribution $\rho$ of DM. As the bin widths are small, for each energy bin 
$[E_{i-1/2},E_{i+1/2}]$, we approximate $dN_{\gamma}/dE_{\gamma}$ with a constant, $C_i$.
This approximation
enables us to calculate the total log-likelihood of the spectrum $\phi(E)$
from the logarithm of the likelihood map ${\mathcal L}_{ij}$ as
\begin{equation}
\ln{\mathcal L}=\sum_i\ln{\mathcal L}_{ij}|_{\phi_j=\frac{\sigmav}{8{\pi}
m_{\chi}^{2}}\times J\times C_i}.
\end{equation}

We use the standard Fermi Science Tools 
package~\cite{web:FermiTools} 
version v10r0p5 to analyze the Fermi-LAT data. We use the newly released 
Fermi Pass 8 data, with four subsets of different point spread function
(PSF) levels (i.e., PSF0, PSF1, PSF2 and PSF3), recorded from 4 August 2008 
to 4 August 2015. These data are selected from 10$^\circ \times$ 10$^\circ$ 
box regions centered on each dSph, and 500 MeV to 500 GeV energies to 
reduce the impact from the bright Earth limb due to the large PSFs at 
low energies. The events with zenith angles greater than 100$^\circ$ 
are also excluded. These selected data are divided into 100 $\times$ 100 
spatial bins with 0.1$^\circ$ bin size and 24 logarithmically spaced 
energy bins. Using the suggested diffuse background model\footnote{\url{http://fermi.gsfc.nasa.gov/ssc/data/access/lat/BackgroundModels.html}}
including a structured Galactic component and an isotropic component, 
as well as point sources within 15$^\circ$ of each dSph from the third
Fermi catalog (3FGL; \cite{Acero:2015gva}) as astrophysical background, 
we first carry out a standard binned likelihood fitting over the entire 
energy range to get the best-fitting parameters for each point source
and the diffuse components. Then we fix all the parameters of diffuse
backgrounds and known point sources in the ROI, and add a point source at 
the position of the dSph. On varying the flux from the newly added point 
source, we calculate the ${\ln\mathcal L}^{kl}_{ij}$ for the $k$th dSph and 
$l$th subset of data in each energy bin and sum over $l$ to obtain the 
likelihood map ${\mathcal L}^k_{ij}$ for the $k$th dSph.

\subsection{Combination of many dSphs}

\begin{figure*}[!htb]
\includegraphics[width=0.9\textwidth]{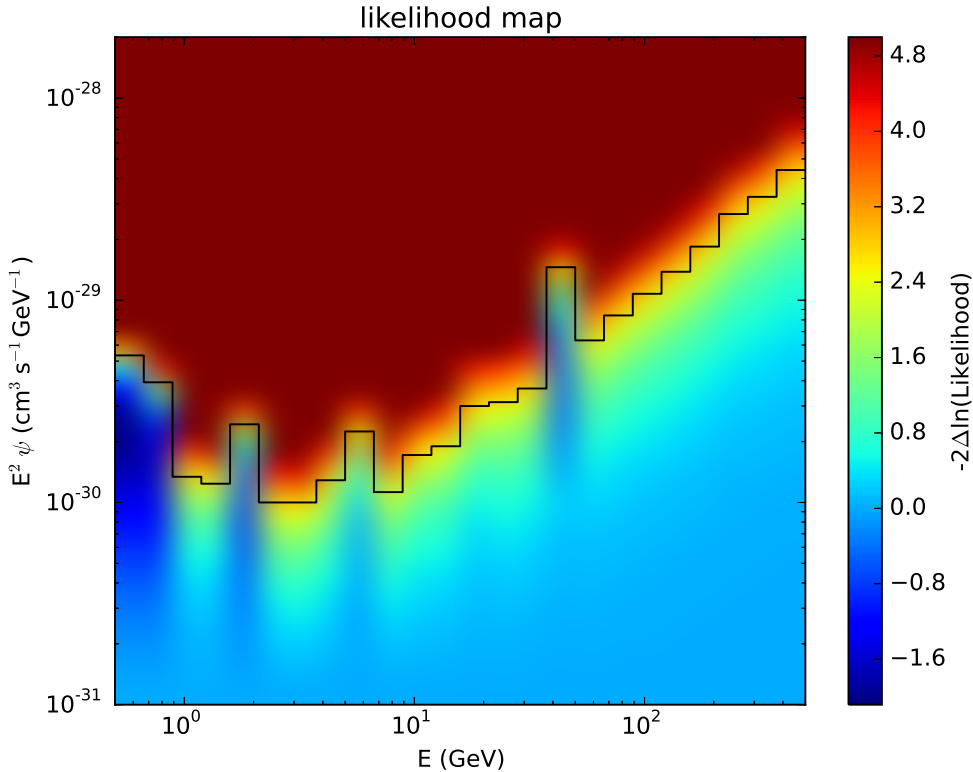}
\caption{The log-likelihood map on the $(E,\,E^2\psi)$ plane based on 7-year 
Fermi-LAT data of the 15 dSphs. The colors show the value of $-2\Delta\ln
{\mathcal L}$, normalized individually for each energy bin (see the text for
details). The region above the solid line is excluded at the 95\% confidence 
level.
\label{fig:lkmp}}
\end{figure*}

If the $J$-factors of dSphs are known, then we can define a new variable, 
$\psi_i=\phi_k(E_i)/J_k=\frac{\sigmav}{8{\pi}m_{\chi}^{2}}\times C_i$,
and derive a combined log-likelihood map on the ($E,\,\psi$) plane by 
adding the log-likelihoods of all dSphs together. Fig. \ref{fig:lkmp}
shows a combined log-likelihood map on the $(E,\,E^2\psi)$ plane, from 
the 15 dSphs as listed in Ref.~\cite{Ackermann:2015zua}. The $J$-factors 
of the dSphs are taken from Ref.~\cite{Bonnivard:2015xpq}.
The solid line shows the one-sided
95\% confidence limit obtained from $-2\Delta\ln{\mathcal L}_i=
-2(\ln{\mathcal L}_i-\ln{\mathcal L}_i^0)=2.71$, where ${\mathcal L}_i^0$
is the likelihood for null-hypothesis (i.e., $\psi_i=0$) for the $i$th
energy bin. For any spectrum $\psi(E)$, the combined log-likelihood can 
be derived via a sum of log-likelihoods in all energy bins.

However, in general the $J$-factors of dSphs cannot be well determined. 
If that is the case, we may not be able to have a combined likelihood map (such as that in Fig. 
\ref{fig:lkmp}) which is independent of $J$-factors\footnote{In Ref. 
\cite{Tsai:2012cs} we profiled $J$-factors in the likelihood function for each 
energy bin, and obtained a combined likelihood map. However, this method multi-counts
the $J$-factor uncertainties. A proper treatment should first combine  
likelihoods in different energy bins and then apply the $J$-factor 
likelihoods \cite{Ackermann:2013yva}.}. We can define a 
likelihood term due to the uncertainties in $J$-factors as 
\cite{Ackermann:2015zua}
\begin{equation}
{\mathcal L}_{J,k}(J_{{\rm obs},k},{\sigma}_k)={\frac{1}{\ln(10)
J_{{\rm obs},k}\sqrt{2\pi}{\sigma}_k}}e^{-[\log_{10}(J_k)-\log_{10}
(J_{{\rm obs},k})]^{2}/{2\sigma_k^2}},
\end{equation}
where $k$ represents the $k$th dSph, $J_k$ is the ``real'' value of the
$J$-factor and $J_{{\rm obs},k}$ is the measured $J$-factor with error 
$\sigma_k$. The joint log-likelihood is then
\begin{equation}
\ln{\mathcal L}({\rm Data}|\phi)={\sum_k}\left({\sum_i}
\ln{\mathcal L}_{ij}|_{\phi_j=\frac{\sigmav}{8{\pi}
m_{\chi}^{2}}\times J\times C_i}+\ln{\mathcal L}_{J,k}\right).
\end{equation}
Maximizing the above joint log-likelihood by varying $J_k$ for each 
dSph, we can obtain the final log-likelihood of the spectrum $\phi(E)$. 

\begin{figure*}[!htb]
\includegraphics[width=0.9\textwidth]{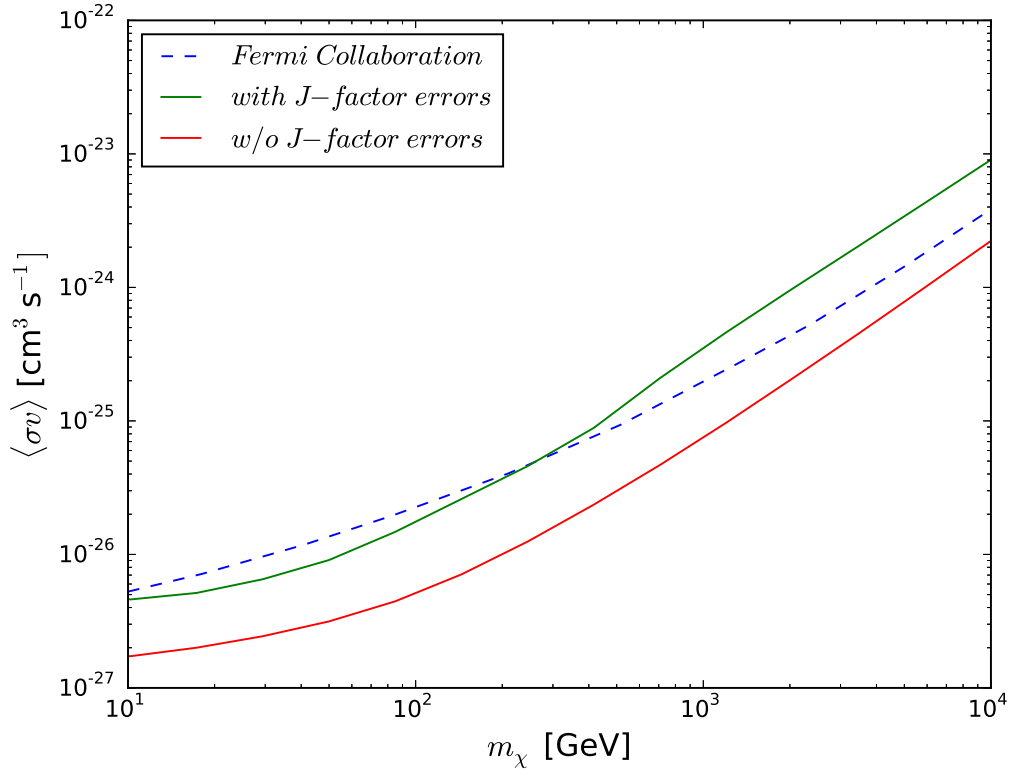}
\caption{95\% upper limits on the DM annihilation cross section for the 
$b\bar{b}$ annihilation channel, derived from a combined analysis of 
Fermi-LAT observations of 15 dSphs. The result obtained by Fermi-LAT
collaboration with 6 year Pass 8 data is shown for comparison
\cite{Ackermann:2015zua}. The two solid lines show the results with
(green) and without (red) uncertainties of $J$-factor measurements.
\label{fig:dwarf_limit}}
\end{figure*}

In Fig. \ref{fig:dwarf_limit} we show the combined 95\% upper limits 
for the $b\bar{b}$ annihilation channel. Here we adopt the $J$-factors given
in Ref. \cite{Bonnivard:2015xpq}. The two solid lines show the differences
between the cases with (green) and without (red) uncertainties in 
$J$-factor measurements. It shows the potential to improve the 
constraints with better determination of the $J$-factors.

%%#######################################################%%
%\section{Application to specific DM models \label{sec:app}}

\section{Description of the code}
\label{sec:manual}

In this section we describe the structure of the \likedm\ code. Users can 
download the source code from Ref.~\cite{codeurl} or the batch file from
the ancillary files to this paper on the arXiv website. \likedm\ is written 
in \texttt{Fortran95}, with a \texttt{Python} interface.

\subsection{Installation}
\likedm\ uses the external package \texttt{Minuit}~\cite{James:1975dr} to 
maximize the likelihoods, which needs to be installed first. To install 
pyLikeDM, the ``\texttt{f2py}'' package is required. We provide a \texttt{BASH} 
script (\url{create_LikeDM.sh}) for quick installation. 
After running \url{create_LikeDM.sh}, the user is prompted to enter a method of \texttt{pyminuit} installation:
\begin{BASH} 
./create_LikeDM.sh 

+++++++++++++++++++++++++++++++++++++++++++++++++++++++++++++++++
Start installing pyminuit
+++++++++++++++++++++++++++++++++++++++++++++++++++++++++++++++++
# Two ways to install pyminuit: 
# (enter "use_pip" or "local" and other keys for doing nothing.)
local
.
.
.
End installing pyLikeDM
Enjoy use!
\end{BASH}
%Either the compiler \texttt{ifort} or \texttt{gfortran} can be chosen.
%{\it However, it seems that the \texttt{Fortran} version of \texttt{Minuit} 
%is not able to run stably with \texttt{gfortran}. Therefore we strongly 
%recommend to use \texttt{ifort} for those who want to run \likedm\ with
%the \texttt{Fortran} interface. Otherwise, the use is better to use 
%the \texttt{Python} interface if only \texttt{gfortran} is available.}

There are three options: \url{use_pip}, \texttt{local}, and any other 
key. If one chooses \url{use_pip}, \texttt{sudo} 
authorization is required to install \texttt{iminuit} via \texttt{pip}. 
If the user does not have \texttt{pip} installed, he/she can install 
\texttt{pyminuit} by using the \texttt{local} option. This step can be
skipped if \texttt{pyminuit} has already been installed.

\likedm\ has been successfully installed and tested under \texttt{Scientific 
Linux}, \texttt{Fedora}, and \texttt{Ubuntu} operating systems.

\subsection{Running \likedm}

The \likedm\ code can be called by
\begin{BASH} 
./pyLikeDM.py LikeDM_input_example.ini [dnde.spec]
\end{BASH}
%or in the \texttt{Fortran} interface as
%\begin{BASH} 
%./LikeDM.fortran LikeDM_input_example.ini [dnde.spec]
%\end{BASH}
where \texttt{LikeDM\_input\_example.ini} is an example file of the input
parameters (see below part C for details), and the argument \texttt{dnde.spec}
is optional, depending on the value of the logical parameter 
\texttt{use\_pppc4}. If \texttt{use\_pppc4=T}, then the DM annihilation
or decay yield spectrum $dN/dE$ is computed using the \texttt{PPPC4} tables
\cite{Cirelli:2010xx}. Otherwise, the file \texttt{dnde.spec} with the
spectrum generated by the user needs to be provided. The output looks like
\begin{BASH} 

  LikeDM (version 1.0)
 
 ************************************************************
 --------- dSphs result: delta [chisq/-2ln(likelihood)] >>>>>
  Fermi_dSphs:   1.3990486012771726     
 --------- dSphs result: delta [chisq/-2ln(likelihood)] <<<<<
 ************************************************************
 
 ************************************************************
 --------- charged particle result >>>>>
 AMS02efr:   41.720980097527232     
 AMS02e+:   27.994316562248855     
 AMS02e-:   33.345732360973315     
 AMS02e+e-:   28.933019605100434     
 AMS02_total_ep:   131.99404862584984     
 
 PAMELA_pbar:   11.168610838448940     
 --------- charged particle result <<<<<
 ************************************************************
\end{BASH}

\subsection{Inputs and outputs}

We provide an example of the input file, \texttt{LikeDM\_input\_example.ini},
in the main folder of \likedm:
\begin{BASH} 
.
. 
.
output_name= LikeDM2016
# See all the information?
# 0 for chisq results 
# 1 for inputs
# 2 for fitting (alpha,beta)
# 3 for input dNdE
# 4 for propagated fluxes of e+ and pbar
# 5 for individual dSph spectrum
# >=6 for fitting results in each step, very slow!
seebug=0  #debug_level

# Which gamma-ray likelihood MAP you are going to include?
# (The way to generate likelihood map can be found in arxiv 1212.3990)
# optimal likelihood map for annihilation DM
dsphs_map=./dat/GaLikeMap/likelihood_fix_p8_psf0123.dat
# optimal likelihood map for decaying DM
#dsphs_map=./dat/GaLikeMap/likelihood_ext_psf0123.dat

#solor modulation potential
epmod=0.6  #GV, positron
apmod=0.6  #GV, antiproton

# What is the DM halo you want to use during propagation? 
WhatHalo=1  #WhatHalo, 1 for NFW, 2 for Einasto and 3 for isothermal

# 6 propagation parameters combination 
# See propagation model in 1205.6474.
# 1-6 correspond to Table I from left to right.
WhatGALPROP=2  #propagation parameters combination


use_dSphs=T        # use_dSphs
use_ep=T           # use_ep
use_ap=T           # use_ap


# If users want to compute decaying DM, this flag should be True.
# Then, code will read decay_time instead of sigmav. 
decayDM=F

DMmass=104.00
sigmav=1e-26 # sigma v [cm^3/s] for annihilation 
decay_time=1e26 # tau [s] for decay

#  T : use PPPC4 Table 
#  False : use external Table from 2nd arguement, ./LikeDM.exe LikeDMexample.ini dnde.txt   
use_pppc4= T       # use PPPC4 dnde

# Branch ratio of xx-> SM SM or x -> SM SM
BR_3=0.0 # e
BR_6=0.0 # mu
BR_9=0.0 # tau
BR_12=1.0 # b
BR_16=0.0 # W
BR_19=0.0 # Z

\end{BASH}
This input format is exactly the same as that in \texttt{CosmoMC}
\cite{Lewis:2002ah} and \texttt{SuperBayes}~\cite{deAustri:2006jwj}.
%The modules used to read the input file are \texttt{src/inifile.f90} 
%for the \texttt{Fortran} interface and \texttt{pyCode/Read\_parameters.py} 
%for the \texttt{Python} interface. 
The modules to read the input file are \texttt{src/Read\_parameters.py} 
and alternatively \texttt{src/inifile.f90}. The parameters are explained 
below:

\begin{itemize}
 \item \texttt{output\_name} \\
 The name of the prefix of the output files. 
 For the Python interface, this is not used because the output is shown on the screen. 
 However, a user can always modify the subroutine \texttt{print\_debug\_info} to store 
 the output with the name defined by this flag.  
 
 \item \texttt{seebug}\\
 An integer number to control the debug information shown on the screen.
 \texttt{seebug=0} to 6 will print different kinds of results, for
 debugging or any interesting outputs such as the pre- or post-propagated
 particle spectra, and the fitting results of nuisance parameters. 
 
 \item \texttt{dsphs\_map}\\
 Likelihood map of dSphs. The full path of the map is needed. 
 For the case of decaying DM, the optimal likelihood map is recommended.     
  
 \item \texttt{epmod} and \texttt{apmod}\\
 Solar modulation potentials in units of GV, for electrons/positrons and 
 protons/antiprotons, respectively.
 
 \item \texttt{WhatHalo}\\
 An integer number to specify the DM halo profile. 1 for NFW, 2 for 
 Einasto and 3 for isothermal. 
    
 \item \texttt{WhatGALPROP}\\
 An integer number to determine the propagation parameters.
 1 to 6 corresponds to the six sets of propagation parameters given in
 Ref.~\cite{Ackermann:2012rg} (see also Table \ref{table:prop} of this
 manual).
 
 \item \texttt{use\_dSphs}, \texttt{use\_ep}, and \texttt{use\_ap}\\
 Logical flags to choose whether or not to use the corresponding data.
 The current version includes Fermi $\gamma$-ray data from dSphs, AMS-02
 $e^+e^-$ data, and PAMELA $\bar{p}$ data.
  
 \item \texttt{decayDM} \\
 Logical flag to determine whether the DM annihilates or decays.

 \item \texttt{DMmass}, \texttt{sigmav}, \texttt{decay\_time} \\
 The DM mass in GeV, annihilation cross section in cm$^3$s$^{-1}$, and 
 decaying lifetime in s, respectively. \texttt{sigmav} takes effect when \texttt{decayDM=F}, 
 and \texttt{decay\_time} takes effects when \texttt{decayDM=T}.
 
 \item \texttt{use\_pppc4} \\
 Logical flag to specify whether to use the PPPC4 table to calculate 
 $dN/dE$. If \texttt{F}, an external file needs to be provided by the user.
 {\it The file needs to be 4 columns, with $E$ in GeV, $\left.\frac{dN}{dE}
 \right|_{\gamma}$ in GeV$^{-1}$, $\left.\frac{dN}{dE}\right|_{e^+}$ in 
 GeV$^{-1}$, and $\left.\frac{dN}{dE}\right|_{\bar{p}}$ in GeV$^{-1}$, 
 respectively.}
 
 \item \texttt{BR\_x} \\
 Branching ratios for different channels when using the PPPC4 table.
 The identification numbers can be found either at the \texttt{PPPC4} website 
 or in the beginning of the file \texttt{src/PYTHIA\_PPPC4.f90}.
  
\end{itemize} 

The outputs include the computed $\chi^2$ values on the screen. Users 
can easily modify the code \texttt{src/monitorLikeDM.f90} to generate
their own favored outputs or store the outputs to a file.
%probably two \texttt{output\_namexxx.mn.out} files which contain the
%fitting results of the nuisance parameters $\alpha_i$ and $\beta_i$
%by \texttt{Minuit}.

\subsection{Package roadmap}
The source code of \likedm\ is located in the \texttt{src/} folder.
%The main routine is \texttt{Main.f90} for the \texttt{Fortran} interface
%or \texttt{pyLikeDM.py} for the \texttt{Python} interface. We introduce 
The main routine is \texttt{pyLikeDM.py} for the \texttt{Python} interface. We introduce 
the other routines grouped by their functionality:

\begin{itemize}

\item \underline{Initialization and Reading tables} \\
\begin{lstlisting}[backgroundcolor=\color{cyan},frame=single,basicstyle=\ttfamily]
src/ReadTable.f90
src/PYTHIA_PPPC4.f90
src/inifile.f90
\end{lstlisting}

The routine \texttt{src/ReadTable.f90} reads the tables of the dSph likelihood
map, the Green's functions for the propagation of positrons and antiprotons
in the Galaxy, and the DM annihilation/decay spectra $dN/dE$ either from
PPPC4 (connected with \texttt{src/PYTHIA\_PPPC4.f90}) or the user supplied
external file. 

In addition to reading the tables, we also collect all the initialization 
subroutines in the \texttt{src/ReadTable.f90} module and hence 
this module is the heart of \likedm.  

The module \texttt{src/inifile.f90} is taken from \texttt{CosmoMC}. It reads the parameter file and sets default values of the parameters.
It is \emph{not} used by default but a user can use this module 
if they wish to construct their own interface.

\item \underline{Gamma-rays from dSphs} \\
\begin{lstlisting}[backgroundcolor=\color{cyan},frame=single,basicstyle=\ttfamily]
src/gamma_dSphs.f90
\end{lstlisting}

This module provides the computation of DM annihilation/decay fluxes from
a set of dSphs and their combined likelihood. The $J$-factors of these
dSphs have been implemented in the likelihood calculation with a profile
likelihood method. By default, a total of 15 dSphs, which are  
Bootes I, Canes Venatici II, Carina, Coma, Draco, Fornax, 
Hercules, Leo II, Leo IV, Sculptor, Segue I, Sextans, Ursa Major II, 
Ursa Minor, and Willman I, are included in the current version
of \likedm. 
Users can enable or disable some dSphs likelihood by turning on/off the flags 
\texttt{dsphs\_\_use} in the \texttt{src/ReadTable.f90} module. 
The $J$-factors are taken from Ref. \cite{Bonnivard:2015xpq} 
for both annihilating and decaying DM.
 
\item \underline{Charged cosmic rays: background} \\
\begin{lstlisting}[backgroundcolor=\color{cyan},frame=single,basicstyle=\ttfamily]
src/charge_bkg.f90
\end{lstlisting}

This routine calculates the background fluxes of $e^+e^-$ and $\bar{p}$
using the empirical formulae described in Sec. II-C. 

\item \underline{Charged cosmic ray: DM $e^+$ and $\bar{p}$} \\
\begin{lstlisting}[backgroundcolor=\color{cyan},frame=single,basicstyle=\ttfamily]
src/charge_lepton.f90
src/charge_antip.f90
\end{lstlisting}

These two routines compute the propagated fluxes of positrons and antiprotons
from DM annihilation or decay, using the Green's function method described
in Sec. II-B. 
%These nuisance parameters at Eq.~\ref{eq:nu_ep},~\ref{eq:nu_em}, and 
%\ref{eq:nu_ap} can be adjusted at the subroutine \texttt{adjustShape\_epem} 
%and \texttt{adjustShape\_ap} but the solar modulation is also included in 
%these two subroutines by calling the subroutine \texttt{AfterModul}. Notes 
%that the DM fluxes outputs from subroutine \texttt{output\_ep\_em\_atEarth} 
%and \texttt{output\_ap\_atEarth} are always included the background and ready 
%to compare with the data. If users want output only the DM component, they 
%can call \texttt{DM\_signal\_epem} and \texttt{DM\_signal\_ap} with the 
%given parameters, 
%\begin{description}
%\item[\texttt{mx}]: input, DM mass
%\item[\texttt{sv}]: input, DM cross section in unit $cm^{-3} s^{-1}$ for 
%annihilation or inverse decay time $\tau^{-1}$ in the unit $s^{-1}$.  
%\item[\texttt{npts}]: input, the length of reference energy and resulting 
%fluxes array described in below.
%\item[\texttt{Ein}]: input, the reference energy array, users' interesting 
%charged particle energy at the earth. 
%\item[\texttt{fluxout}]: output, the DM fluxes with length \texttt{npts}, 
%by given information of above parameters, propagation Green function but 
%the solar modulation and reshaped effect not included.
%\end{description}

\item \underline{Charged cosmic rays: datasets} \\
\begin{lstlisting}[backgroundcolor=\color{cyan},frame=single,basicstyle=\ttfamily]
src/charge_data.f90
\end{lstlisting}
  
This routine gives the cosmic ray data from AMS-02 \cite{Aguilar:2013qda,
Accardo:2014lma,Aguilar:2014mma,Aguilar:2014fea} and PAMELA 
\cite{Adriani:2010rc}, and returns the calculated $\chi^2$ values for
given theoretical fluxes.

\item \underline{Auxiliary module}\\

\begin{lstlisting}[backgroundcolor=\color{cyan},frame=single,basicstyle=\ttfamily]
src/MathLib.f90 
src/monitorLikeDM.f90
src/Main.f90 
\end{lstlisting}

The file \texttt{src/MathLib.f90} provides some useful mathematical tools 
such as interpolation and integration. 
The routine \texttt{src/monitorLikeDM.f90} gives the outputs controlled 
by the flag \texttt{seebug}. We also have a main routine, 
\texttt{src/Main.f90}, which is currently \emph{not} used in the \texttt{Python} 
version but left as an alternative in the pure \texttt{Fortran} version.

%Finally, the module \texttt{PYTHIA\_PPPC4} is used to interface the 
%\texttt{PPPC4} pythia tables. We copy some relevant information from 
%\texttt{PPPC4} website and paste them in the header of this module. 
%Users should read them before use it. The \texttt{PPPC4} pythia tables 
%will be stored in the following parameters at the beginning of this module. 

%\begin{F95} 
%    integer :: PPPC4_Mbin
%    integer :: PPPC4_Ebin
%    real*8 :: PPPC4_spect(3,62,179)   ! dnde from PPPC4 table products, [mx bins, energy bins].
%    real*8 :: PPPC4_K_over_mx(179)
%    real*8 :: PPPC4_mx(62)
%    real*8 :: PPPC4_dnde(4,179)       ! dnde for LikeDM formate
%\end{F95} 

%The DM mass bin number is 62 (\texttt{PPPC4\_Mbin$=62$}) and 
%the spectrum energy bin number is 179 (\texttt{PPPC4\_Ebin$=179$}). 
%All the tables are originally from PPPC4. 
%It has to note that those table do not use for DM spectrum computation. 
%Hence, if users want to use them, \texttt{PPPC4\_dnde} has to be copied to 
%table \texttt{PYTHIA\_\_dnde} like the below example,
%\begin{F95}
%      PYTHIA__nrows=179
%      PYTHIA__dnde(1,1:PYTHIA__nrows)=PPPC4_dnde(1,1:PYTHIA__nrows)
%      PYTHIA__dnde(2,1:PYTHIA__nrows)=PPPC4_dnde(2,1:PYTHIA__nrows)
%      PYTHIA__dnde(3,1:PYTHIA__nrows)=PPPC4_dnde(3,1:PYTHIA__nrows)
%      PYTHIA__dnde(4,1:PYTHIA__nrows)=PPPC4_dnde(4,1:PYTHIA__nrows)
%\end{F95} 
%Users can find the above lines at the subroutine \texttt{Load\_dnde} 
%of module \texttt{ReadTable}.         

\end{itemize}

\section{Summary \label{sec:sum}}

We present a publicly-available tool, \likedm, for likelihood
calculations in DM models. It enables fast
computation of the likelihood of a given DM model (defined by mass, cross section
or decay rate, and annihilation or decay yield spectrum), without digging into
the details of CR propagation, Fermi-LAT data analysis, or related 
astrophysical backgrounds. This code depends only on the \texttt{Minuit}
minimization package, and is easy to install and run. 
The code \likedm\ also provides an easy framework that can be linked to any particle 
model or Monte-Carlo code to perform a global study.

The currently released version (v1.0) contains only the indirect detection data, 
including the electron/positron measurements by AMS-02, the antiproton 
measurements by PAMELA, and the $\gamma$-ray observations from dSphs by 
Fermi-LAT. Further developments with more data, e.g., from the $\gamma$-ray 
observations of the Galactic center and isotropic background, as well
as underground direct detection data, will be carried out soon.

\section*{Acknowledgments}
We thank Vincent Bonnivard who kindly provides the MCMC results about 
the profile parameters of dSphs, and Andrew Fowlie and Shankha Banerjee 
who carefully read and improve the presentation of the manuscript. 
Y.S.T. was supported by World Premier 
International Research Center Initiative (WPI), MEXT, Japan. 
Q.Y. was supported by the National Key Program for Research and Development (No. 2016YFA0400200) 
and the 100 Talents program of Chinese Academy of Sciences.

%%%%%%%%%%%%%%%%%%%%%%%%%%%%%%%%%%%%%%%%%%%%%%%%%%%%



%% References


\bibliographystyle{unsrt}

\end{document}